\documentclass[a4paper]{jpconf}
\pdfoutput=1

\usepackage[T1]{fontenc} 
\usepackage{color}
\usepackage[english]{babel} 
\usepackage{amsmath,amsfonts,amsthm,amssymb} 
\usepackage{graphicx,sidecap,caption}
\usepackage{hyperref,cite}
\allowdisplaybreaks

\numberwithin{equation}{section} 
\numberwithin{figure}{section}
\numberwithin{table}{section} 
\begin{document}

\title{Primordial GWs from universality classes of pseudo-scalar inflation}
\author{M. Pieroni}
 \address{Laboratoire AstroParticule et Cosmologie (APC),
Universit\'e Paris Diderot}
\address{Paris Centre for Cosmological Physics (PCCP), F75205 Paris Cedex 13}
\ead{mauropieroni@gmail.com, mauro.pieroni@apc.in2p3.fr}
\begin{abstract}
In this contribution we discuss the possibility of generating an observable gravitational wave (GW) background by coupling a pseudo-scalar inflaton to some Abelian gauge fields. This analysis is performed by dividing inflationary models into universality classes. We find that of the most promising scenario is a Starobinsky-like model, which may lead to the generation of observational signatures both in upcoming CMB detectors as well as for direct GW detectors. The signal which can be produced in these models would both be observable in ground-based detectors, such as advanced LIGO, and in space-based detectors, such as LISA. The complementarity between the CMB and direct GW detection may be used to extract informations on the microphysics of inflation. Interestingly the mechanism discussed in this contribution may also be relevant for the generation of Primordial Black Holes (PBHs).
\end{abstract}

\section{Introduction}
\label{introduction}
Inflation is nowadays accepted as a cornerstone of modern cosmology and in particular its simplest realization in terms of single scalar field slow-rolling in its potential appears to match with cosmological observations at CMB scales. During inflation, quantum fluctuations
of the inflaton field and of the metric are stretched on macroscopic scales by the exponential expansion. This mechanism naturally provides an explanation to the anisotropies in the CMB and it is also expected to generate a background of primordial GWs. Because of the weakness of their interactions, GWs are practically freely traveling through the Universe and thus they can provide precise information on times as early as inflation. As a consequence the detection of this primordial GW background at direct GW detectors would deepen our understanding of the physics of the early Universe. Unfortunately, the spectrum of GWs produced during slow-roll inflation lies below the range of current and upcoming direct GW detectors.

While this minimal setup is both simple and effective, it is not completely satisfactory from a theoretical point of view. For this reason, several generalized landscape (such as multifields models or models with non-standard kinetic terms) have been proposed in order to define a realistic model of inflation that is also consistent with our knowledge of the fundamental interactions. In this contribution we discuss a realization of inflation where the inflaton $\phi$ (which is considered to be a pseudo-scalar) is coupled to some Abelian gauge fields through a generic higher-dimensional coupling~\cite{Cook:2011hg} $\sim \phi \, F_{\mu\nu} \, \tilde{F}^{\mu\nu}$ (where $F_{\mu\nu}$ is the usual field strength tensor and $\tilde{F}^{\mu\nu}$ is the dual field strength tensor). Remarkably, this term generates an instability, which leads to an exponential production of the gauge fields~\cite{Turner:1987bw, Garretson:1992vt, Anber:2006xt}. As a consequence, the gauge fields produced with this mechanism induce a back-reaction both on the background dynamics~\cite{Anber:2009ua,Barnaby} and on the perturbations~\cite{Anber:2012du,Linde:2012bt} (both the scalar and tensor power spectra are exponentially enhanced at small scales). 

This generalized framework leads to a wide set of observational consequences such as: the production of a GW background which can be observed at GW interferometers such as advanced LIGO/VIRGO and LISA~\cite{Cook:2011hg,Barnaby,Domcke:2016bkh,Pieroni:2016gdg}, the presence of a non-Gaussian component in the scalar power spectrum~\cite{Barnaby,Anber:2012du,Barnaby:2010vf,Linde:2012bt,Domcke:2016bkh,Pieroni:2016gdg}, the generation of a distribution of PBHs~\cite{Linde:2012bt,Domcke:2016bkh,Pieroni:2016gdg} and the generation of $\mu$-distortions~\cite{Meerburg:2012id,Domcke:2016bkh,Pieroni:2016gdg}. 

In this contribution, we discuss the shape of the scalar and tensor spectra that are generated in this generalized framework. In particular we show that because of the exponential enhancement of the tensor spectrum at small scales, GW interferometers may be used to probe different inflationary models. The analysis presented in this contribution is carried out by grouping models into universality classes~\cite{Pieroni:2016gdg,Classification}. As this characterization is based only on the asymptotic properties of the potential, it leads to a more general description of the problem. The results presented in this proceeding are based on work with Valerie Domcke and Pierre Bin\'etruy~\cite{Domcke:2016bkh}.
 
\section{Pseudo-scalar inflation in presence of gauge fields.}
\label{pseudoscalar}
We consider the action~\cite{Turner:1987bw, Garretson:1992vt, Anber:2006xt,Anber:2012du,Anber:2009ua,Barnaby,Linde:2012bt} for a pseudo-scalar inflaton $\phi$ coupled to a certain number $\mathcal{N}$ of Abelian gauge fields $A_\mu^a$ associated to some $U(1)$ gauge symmetries:
\begin{equation}
\label{action_pseudoscalar}
\mathcal{S}= \int \textrm{d}^4 x \sqrt{|g|} \left[m_p^2 \frac{R}{2} -\frac{1}{2} \partial_\mu \phi \partial^\mu \phi - V(\phi) - \frac{1}{4} F^a_{\mu \nu} F_a^{\mu \nu} - \frac{\alpha^a}{4 \Lambda} \phi F^a_{\mu \nu} \tilde{F}_a^{\mu \nu} \right ]\ ,
\end{equation}
where $m_p \simeq 2.4 \cdot 10^{18}\,$~GeV denotes the reduced Planck mass, $V(\phi)$ is the scalar potential, $F^a_{\mu \nu}$ is the usual field-strength tensor, $\tilde{F}_a^{\mu \nu}$ is the dual field-strength tensor, $\Lambda$ is a mass scale (that suppresses the higher-dimensional scalar-vector coupling) and $\alpha^a$ are the dimensionless coupling constants of the Abelian gauge fields. In the following we consider $\alpha^a = \alpha$ for all $a = \{1,2,..\mathcal{N}\}$. The equations of motion which describe the classical evolution of $\phi$ and $\vec{A}^a$ read: 
\begin{align}
\label{eq_motion}
\ddot \phi + 3 H \dot{\phi} + \frac{\partial V}{\partial \phi}  = \frac{\alpha}{\Lambda} \langle \vec{E}^a \cdot \vec{B}^a \rangle \ , \qquad  \frac{d^2}{d \tau^2}\vec{A}^a - \nabla^2 \vec{A}^a - \frac{\alpha}{\Lambda} \frac{\textrm{d} \phi}{\textrm{d} \tau} \nabla \times \vec{A}^a  = 0 \ ,
\end{align}
where dots are used to denote derivatives with respect to cosmic time $t$, $\tau$ is the conformal time defined as $\textrm{d}\tau = \textrm{d}t/a(t)$, the brackets $\langle \cdot \rangle$ denote a spatial mean and the vectors $\vec{E}^a$ and $\vec{B}^a$ are the ``electric'' and ``magnetic'' fields defined as:
\begin{equation}
  \label{electric_magnetic}
  \vec{E}^a \equiv -\frac{1 }{a^2} \frac{\textrm{d} \vec{A}^a}{\textrm{d} \tau} = -\frac{1 }{a} \frac{\textrm{d} \vec{A}^a}{\textrm{d} t} \ , \qquad \qquad  \vec{B}^a \equiv \frac{1}{a^2} \vec{\nabla} \times \vec{A}^a  \ .
\end{equation}
As we show in the following, assuming $\dot{ \phi} $ to be slowly varying (that indeed is satisfied during slow-roll inflation), the equation of motion for $\vec{A}^a$ can be solved analytically. Once this solution is found, we can plug it into the equation of motion for $\phi $ and study the back-reaction. 
By performing a spatial Fourier transform, the equation of motion for the gauge fields reads:
\begin{equation}
\label{eq_motionAfourier}
  \frac{\textrm{d}^2 \ \tilde{A}^{a}_{\pm}(\tau,\vec{k})}{\textrm{d} \tau^2}  + \left[ k^2 \pm 2k  \frac{\xi}{\tau} \right]\tilde{A}^{a}_{\pm}(\tau,\vec{k}) =  \ 0 \ , 
  \end{equation}
where, given the two helicity modes $\hat{e}_{\pm}$, we have defined $\tilde{A}^a_{\pm}(\tau,\vec{k})$ as $\tilde{\vec{A}}^{a}_{\pm} = \hat{e}_{\pm} \tilde{A}^{a}_{\pm}  \exp(i \vec{k}\cdot \vec{x})$ and we have introduced the dimensionless parameter $\xi$ defined by $\xi \equiv \alpha |\dot{\phi}|/(2 \Lambda H) $. For slowly varying $\xi$ this equation describes a tachyonic instability in the $A_+$ mode that induces an exponential growth for modes at low $k$. In particular, the mode $\tilde{A}_+$ can be expressed as~\cite{Anber:2009ua,Barnaby:2010vf,Barnaby}:
\begin{equation}
\tilde{A}_+^a \simeq \frac{1}{\sqrt{2k}} \left( \frac{k}{2 \xi a H}\right)^{1/4} e^{ \pi \xi - 2 \sqrt{2 \xi k/(a H)}} \ .
\end{equation}
At this point it is interesting to notice that $\xi\sim \dot{\phi}/H\sim \sqrt{\epsilon_H}$ (where $\epsilon_H$ is the first slow-roll parameter $\epsilon_H \equiv -\dot{H}/H^2\simeq \epsilon_V \equiv (V_{,\phi}/V)^2 /(2 m_p^2) $). As we have $\epsilon_H \ll 1$ at CMB scales and $\epsilon_H = 1$ at the end of inflation, we may have a negligible gauge field production at CMB scales (consistently with CMB observations) and a strong production towards the end of inflation. 

The back-reaction on the equation of motion for $\phi$ is given by $\langle \vec{E}^a \cdot \vec{B}^a \rangle = \mathcal{N} \cdot \, 2.4 \cdot 10^{-4} H^4 /\xi^4 e^{2 \pi \xi} $. This term plays the role of an additional friction term that does not effect the evolution at CMB scales (where $\xi \propto \sqrt{\epsilon_H} \ll 1$) but it strongly affects the last part of the evolution. It is important to stress that, by modifying the background dynamics, this mechanism affects the standard inflationary predictions~\cite{Domcke:2016bkh,Pieroni:2016gdg}. In particular the predicted value of scalar spectral index $n_s$ gets increased and the predicted value of the tensor-to-scalar ration $r$ gets decreased with respect to the case with $\alpha/\Lambda = 0$.

As already mentioned in Sec.~\ref{introduction}, the strong production of gauge fields is not only affecting the background evolution but it is also modifying the scalar and tensor power spectra. In particular, the presence of the gauge fields induces a source term in the equation of motion for perturbations leading to an additional contribution to the spectra. 

Let us start by considering the expression for the scalar power spectrum~\cite{Linde:2012bt,Domcke:2016bkh}:
\begin{equation}
\Delta^2_s(k) = \Delta^2_s(k)_\text{vac} + \Delta^2_s(k)_\text{gauge} = \left(\frac{H^2}{2 \pi |\dot{\phi}|}\right)^2 + \left( \frac{\alpha \langle \vec{E}^a\cdot \vec{B}^a \rangle/ \sqrt\mathcal{N}}{3 \Lambda b H \dot{\phi}} \right)^2 .
\label{scalar}
\end{equation}
where we have defined $b \equiv 1 - 2 \pi \xi \alpha \langle \vec{E}^a \cdot \vec{B}^a \rangle / (3 \Lambda H \dot{\phi}) $. As $\langle \vec{E}^a \cdot \vec{B}^a \rangle \simeq \mathcal{N} \cdot \langle \vec{E} \cdot \vec{B} \rangle$ ($\langle \vec{E} \cdot \vec{B} \rangle$ is the value of $\langle \vec{E}^a \cdot \vec{B}^a \rangle$ for $\mathcal{N} =1$), it is possible to show that for large values of $\xi$ (\emph{i.e.} at small scales) we get: 
\begin{equation}
  \label{scalar_strong}
  \Delta^2_s(k) \simeq \frac{1}{\mathcal{N} (2 \pi \xi)^2} \ .
\end{equation}
Interestingly, with a proper choice of the parameters of the model, we recover the usual scale-invariant power spectrum of inflation at large scales (\emph{i.e.} scales probed through CMB observations) and a nearly flat spectrum at small scales. The strong amplification of the scalar spectrum at small scales may lead to the generation of a distribution of massive PBHs~\cite{Linde:2012bt}.

The normalized density of GWs at present time can be expressed as~\cite{Barnaby,Domcke:2016bkh}:
\begin{equation}
\Omega_{GW} \equiv \frac{\Omega_{R,0}}{24} \Delta^2_{t} \simeq \frac{1}{12} \Omega_{R,0} \left(  \frac{ H}{ \pi m_p} \right)^2 \left(1 + 4.3 \cdot 10^{-7} \mathcal{N} \frac{H^2}{ m_p^2 \xi^6} e^{4 \pi \xi}\right)\ ,
\label{OmegaGW}
\end{equation}
where $\Omega_{R,0} = 8.6 \cdot 10^{-5}$ denotes the radiation energy density today. Before concluding this section it is crucial to stress that the contribution of one of the two modes of the GWs is suppressed by a $10^{-3}$ factor with respect to the other. As a consequence the new contribution to the GW spectrum is strongly chiral. This is a peculiar characteristic of this model and it offers a method to distinguish this signal from other GW backgrounds.

\section{Universality classes and GWs.}
While in the existing literature~\cite{Turner:1987bw, Garretson:1992vt, Anber:2006xt,Anber:2012du,Anber:2009ua,Barnaby,Linde:2012bt} these models have always been studied by assuming a chaotic potential~\cite{Linde:1983gd} $V(\phi) = m^2 \phi^2/2$ for the inflaton, in this contribution (following the analysis of~\cite{Domcke:2016bkh}) we extend the analysis to a broader class of models. For this purpose it is useful to notice that a large set of single-field slow-roll models can be expressed in terms of the parametrization~\cite{Pieroni:2016gdg,Classification}: 
\begin{equation}
\label{Nparameterization}
\epsilon_H \simeq \epsilon_V \simeq  \frac{\beta_p}{N^p}  + \mathcal{O}(1/N^{p+1})\ ,
\end{equation}
where $N \equiv - \int H \textrm {d}t$ denotes the number of e-foldings from the end of inflation, $\beta_p>0$ and $p>1$ are constants. It is important to stress that using this parametrization we are not completely specifying the potential $V(\phi)$ but only its asymptotic expression. As a single asymptotic behavior can be reached by several models, with this parametrization we can produce an analysis that does not holds for single models but rather for sets of theories that share a single asymptotic behavior \emph{i.e.} for universality classes of models. 

By noticing that the parameter $\xi$ that controls the instability is proportional to the square root of the first slow-roll parameter $\epsilon_H$, we can thus conclude that it is particularly convenient to use the parametrization of Eq.~\eqref{Nparameterization} to produce a general analysis of the model introduced in Sec.~\ref{pseudoscalar}. Indeed (as explained in~\cite{Domcke:2016bkh}) using the parametrization of Eq.~\eqref{Nparameterization} we have direct control on all the parameters that affect the shape of the scalar and tensor spectra \emph{i.e.} $\mathcal{N}$, $\alpha/\Lambda$, $\beta_p$, $p$ and $V_0$ (normalization of the potential). This allows for the definition of a systematic method to scan the parameter space and to identify the different classes of models.

\begin{figure}[htb!]
\centering

{\includegraphics[width=0.49 \columnwidth]{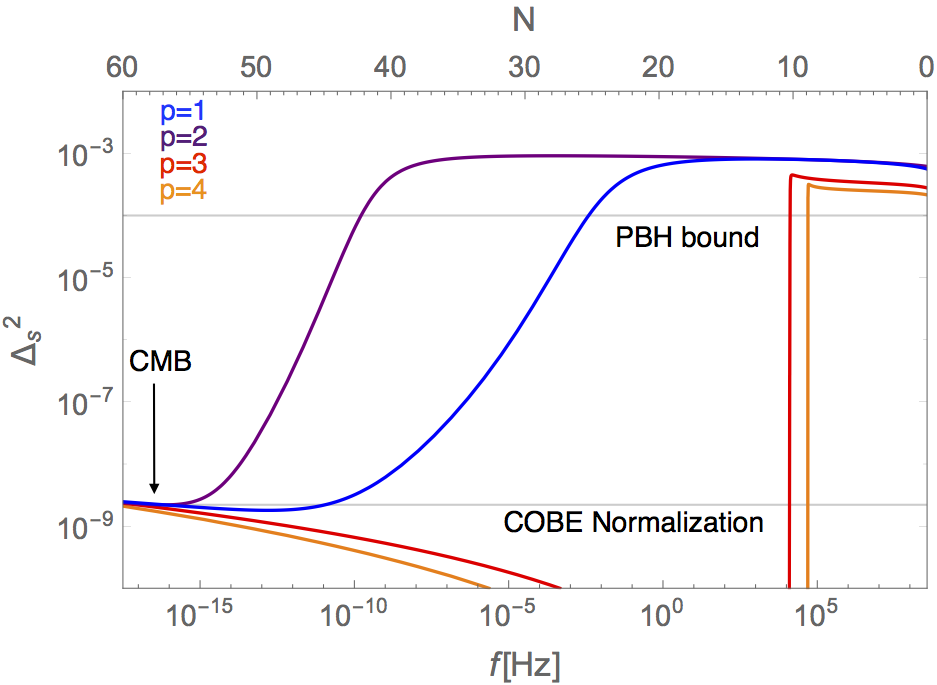}
\includegraphics[width=0.49 \columnwidth]{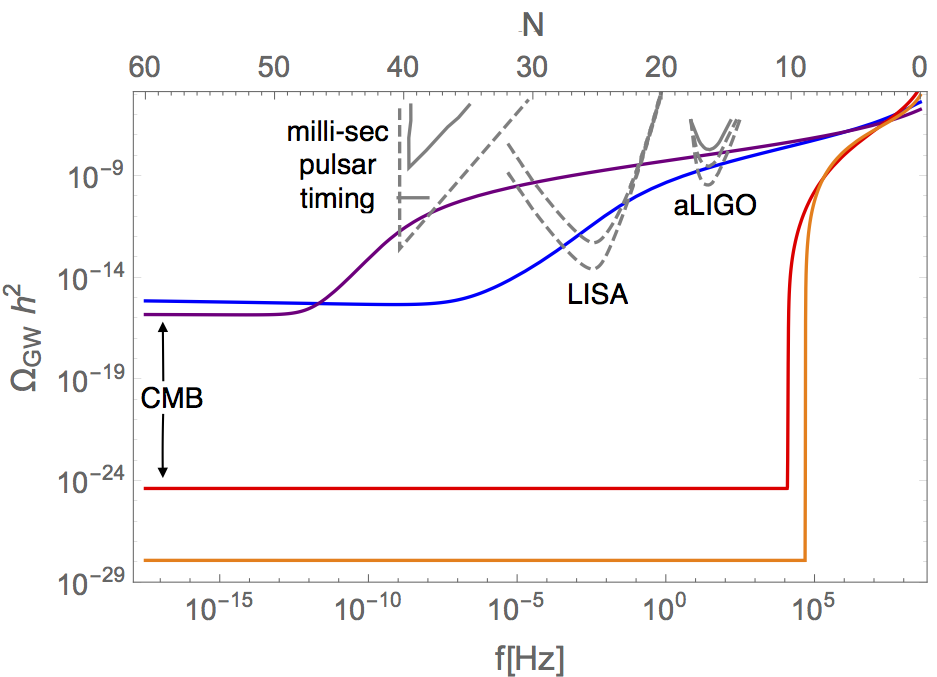}}\\
\caption{\label{figure1} Scalar (left) and tensor (right) power spectra for different universality classes of models with $\mathcal{N}= 1$. The upper horizontal line in the left plot is the estimated PBH bound, the lower horizontal line indicates the COBE normalization. In the plot of the tensor power spectrum we present the sensitivity curves of some present and future direct GW detectors.}
\end{figure}

The models shown in Fig.~\ref{figure1} are: chaotic inflation~\cite{Linde:1983gd} ($p=1$), Starobinsky-like\footnote{Starobinsky-like models of inflation with pseudo-scalar fields can be effectively implemented in the context of supergravity by employing a shift-symmetry in the K\"ahler potential~\cite{Kawasaki:2000yn} (see for example~\cite{SUGRA}).} inflation~\cite{Starobinsky:1980te} ($p=2$) and two implementations of hilltop inflation~\cite{Boubekeur:2005zm} ($p=3,4$). For all the universality classes shown in this plots we have fixed $\mathcal{N}= 1$. Consistently with the $N$-dependence expressed by the parametrization of Eq.~\eqref{Nparameterization}, by increasing the value of $p$ we get a smaller value of $\xi$ at CMB scales and a larger value at the end of inflation (implying a steeper increase). This explains the steeper increase in the spectra for models with large values of $p$ (see Fig.~\ref{figure1}). In order to be consistent with the constraints set by CMB observations~\cite{Ade:2015lrj} the potential $V_0$ is fixed in order to respect the COBE normalization that fixes the amplitude of the scalar spectrum at CMB scales. At a given value for $p$ (which specifies the different universality classes of inflation) we are thus left with only two parameters \emph{i.e.} $\alpha/\Lambda$ and $\beta_p$. The parameter $\alpha/\Lambda$ controls the strength of the interactions between the inflaton and the gauge fields. As a consequence by increasing $\alpha/\Lambda$, the gauge field production gets faster and the enhancement of the spectrum is shifted at earlier times (\emph{i.e.} larger frequencies). Conversely, by varying the parameter $\beta_p$, we change the vacuum amplitude of the tensor spectrum. In particular, by the increasing the value of $\beta_p$ we ``vertically '' shift the GW spectra. The parameters $\alpha/\Lambda$ and $\beta_p$ for the models shown in Fig.~\ref{figure1} are chosen in order to maximize the GW spectrum while obeying the CMB constraints~\cite{Ade:2015lrj}. For example the late rise in both the scalar and tensor spectra for the hilltop models ($p=3,4$), is due to the constraints on $n_s$ scalar spectral index\footnote{For the models described by the parametrization of Eq.~\eqref{Nparameterization} we have $n_s \simeq 1 - p/N$ and thus larger values of $p$ implies smaller values for $n_s$. As already mentioned in Sec.~\ref{pseudoscalar}, the presence of gauge fields effectively reduces the value of $n_s$ and thus CMB measurements~\cite{Ade:2015lrj} set a rather severe constraint on the values of $\alpha/\Lambda$ that are allowed for these models.}.

By observing the plot of the scalar spectra (where the amplitude of the spectrum at CMB scales is fixed) we can see that a larger value for $p$ implies a smaller amplitude at small scales (towards the end of inflation). Such a behavior is consistent with the analytical prediction given in Eq.~\eqref{scalar_strong}. Notice that all the models shown in this plot are in tension (by a $\mathcal{O}(1)$ factor) with the estimate of the PBH bound~\cite{Linde:2012bt}. This discrepancy can be eased by taking into account the theoretical uncertainties in the derivation of the PBH bound. Moreover, by considering models with $\mathcal{N} >1$ it is possible to produce a GW background that can be observed at upcoming direct GW detectors, while respecting this bound~\cite{Domcke:2016bkh,Pieroni:2016gdg}.

For tensor spectra we can see that larger values of $p$ imply a suppression of the amplitude at CMB scales but also a steeper increase of the spectrum towards higher frequencies. It is interesting to 
notice that the tensor spectra are approaching a nearly universal value in the strong gauge field regime. As discussed in~\cite{Domcke:2016bkh}, this behavior can be explained on solid grounds using the condition $\epsilon_H \simeq 1$ at the end of inflation. In particular this condition can be used to impose an absolute upper bound on the GW spectrum $\Omega_{GW}$:
\begin{equation}
\Omega_{GW} h^2 \lesssim 2.4 \cdot 10^{-5} \mathcal{N}^{-1} \ ,
\label{OmegaMax}
\end{equation}
Given the weakness of the assumptions to derive this bound ($\epsilon_H \simeq 1$ at the end of inflation), in practice it can be seen as model independent.

\begin{SCfigure}[][h!] 
\centering
\includegraphics[width=0.49\textwidth]{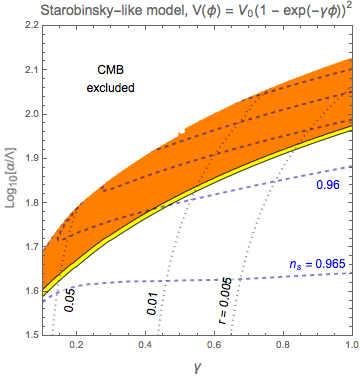} \vspace{0.1cm}
\caption{\label{fig:sidecap} Plot of the $(\gamma, \alpha/\Lambda)$ parameter space for Starobinsky-like models where we have set $\beta_p = 1/(2 \gamma^2)$ (to lighten the notation we have set $m_p =1$ and thus $\Lambda$ is dimensionless). The yellow and orange regions respectively correspond to the regions of the parameter space that can be probed with the A5M5 and A1M2 LISA configurations. Dashed blue lines correspond to fixed values for $n_s$ (from $0.965$ to $0.945$) and dotted gray lines correspond to fixed values of $r$ ($0.05, 0.01, 0.005$). The white region in the top left corner is excluded because of the non-observation of non-gaussianities in the CMB. A similar plot can be produced with the projected sensitivity of the different runs of Advanced LIGO.\\}
\end{SCfigure}

As from the point of view of potential observations Starobinsky-like models ($p=2$) appears to be the most promising among the universality classes shown in Fig.~\ref{figure1}, we proceed by restricting our analysis to this particular class. In particular, in Fig.~\ref{fig:sidecap} we show a scan plot for the parameters of these models that correspond to:
\begin{equation}
   V(\phi) \simeq V_0 \left[ 1 - \exp(\gamma \phi)\right]^2 \ , 
 \end{equation} 
 where $\beta_p = 1/(2 \gamma^2)$. Interestingly this plot shows a complementarity between CMB measurements and GW interferometers. Different measurements are starting to probe the viable parameter space of these models from different directions: while searches for non-Gaussianities in the CMB and direct GW detection probe the region of large $\alpha/\Lambda$, searches for GWs in the CMB constraints the small $\gamma$ region.

\section{Conclusions.}
\label{conclusions}
The general higher-dimensional coupling between a pseudo-scalar inflaton and some Abelian gauge fields induces a tachyonic instability that leads to an exponential production of gauge fields during inflation. The strong production of these gauge fields back-reacts on the system and introduces an additional friction term that affects the last part of inflation. In general this affects the standard inflationary predictions for a given model by decreasing the predicted value of $n_s$ and increasing $r$. The exponential production of gauge fields also leads to a modification of the equation of motion for scalar and tensor perturbations. In particular it gives rise to a strong enhancement of the scalar and tensor power spectra at high frequencies (small scales). This mechanism may lead to the generation of a distribution of massive PBHs which contribute to the fraction of Dark Matter present in the Universe~\cite{PBHs-DM} and to the production of chiral primordial GWs that can be observed at direct GWs detectors. The complementarity between different measurements provide a powerful method to probe these models. Moreover, as direct GWs detectors may access a very wide range of frequencies, they can potentially distinguish between different universality classes. Such a measurement would provide precious information on the microphysics of inflation.

\section*{ Acknowledgements}
We acknowledge the financial support of the UnivEarthS
Labex program at Sorbonne Paris Cit\'e (ANR-10-LABX-0023 and ANR-11-IDEX-0005-02) and the Paris Centre for Cosmological Physics.

\end{document}